\documentclass[11pt,twoside]{article}
\usepackage{newpasp}

\usepackage{epsf}

\index{Peck, A. B.}
\index{Taylor, G. B.}
\index{Menten, K. M.}

\begin{document}

\title{Parsec-Scale Imaging of H\kern0.1em{\sc i}\ Absorption in 1946+708}
\author{A. B. Peck}
\affil{MPIfR, Auf dem H\"ugel 69, D-53121 Bonn, Germany}
\author{G. B. Taylor}
\affil{NRAO, P.O. Box 0, Socorro, NM 87801, USA}
\author{K. M. Menten}
\affil{MPIfR, Auf dem H\"ugel 69, D-53121 Bonn, Germany}


\begin{abstract}

In the last several years, a number of compact extragalactic radio
sources have been found to exhibit neutral hydrogen absorption at or
near the systemic velocities of their host galaxies.  Models proposed
to explain this phenomenon involve a circumnuclear torus of gas and
dust.  The orientation of this structure determines whether or not the
central engine appears obscured.  Understanding the spatial
distribution and kinematics of the H\kern0.1em{\sc i}\ detected toward
the central parsecs of these sources provides an important test of
this model and of unified schemes for AGN.

We present results of Global VLBI Network observations of the
redshifted 21 cm H\kern0.1em{\sc i}\ line toward the Compact Symmetric
Object 1946+708 ($z$=0.101).  This source is of particular interest
because it exhibits bi-directional motion measurable on timescales of
a few years.  The resulting unique information about the geometry of
the continuum source greatly assists in the interpretation of the
H\kern0.1em{\sc i}\ distribution.

We find significant structure in the gas on parsec scales.  The peak
column density of the H\kern0.1em{\sc i}\ occurs near the center of
activity of the source, as does the highest velocity dispersion
(FWHM$\simeq$ 350 to 400 km s$^{-1}$).  The distribution of gas in
1946+708 is strongly suggestive of a circumnuclear torus of atomic
material with one or more additional compact clumps of gas along the
line of sight to the approaching jet.

\end{abstract}




\noindent{\bf 1. Results}

Neutral hydrogen is present toward all of the radio components in the
Compact Symmetric Object 1946+708.  A broad component is seen in all 6
integrated profiles shown in Figure 1.  The width of this broad line varies
significantly across the source, with the broadest linewidth (FWHM
$\sim$375 km s$^{-1}$) occurring near the core of the source.  The
opacity increases gradually toward the receding jet, while the column
density peaks near the core.  A narrow line appears in each profile as
well.  This line is distinguishable from the broad component in
profiles 1 through 4, but is blended with the broad line in profiles 5
\& 6.  The narrow line (FWHM$\sim$30$-$65 km s$^{-1}$) does not vary
much in central velocity.

The systemic velocity obtained from optical observations of both
emission and stellar absorption lines (Stickel \& K\"{u}hr 1993) is
30279$\pm$300 km s$^{-1}$.  Thus all of the H\kern0.1em{\sc i}\ absorption
features reported here are at the systemic velocity to within the
uncertainty of the optical measurements.
\begin{figure}
\vspace{8cm}
\includegraphics{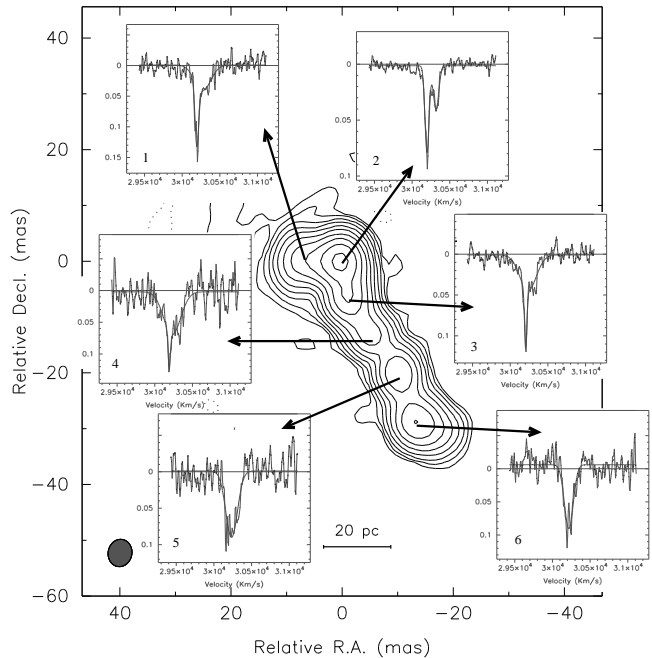}
\caption{\small H\kern0.1em{\sc i}\ absorption profiles toward 1946+708.  Profile 2 shows the absorption toward the approaching jet hotspot, profile 6 -- the receding jet hotspot, and profile 4 is centered on the core of the source.  The beam is 4.3x4.9 mas, and the rms noise in a single channel is $\sim$0.8 mJy/beam.}
\label{fig1}
\end{figure} 

\medskip
\noindent{\bf 2. Conclusions}

The high velocity dispersion and column density toward the core of the
source are indicative of fast moving circumnuclear gas, perhaps in a
rotating toroidal structure.  Further evidence for this region of high
kinetic energy and column density is found in the continuum spectra of
the jet components which indicate a region of free-free absorption
along the line of sight toward the core and inner receding jet (Peck,
Taylor, \& Conway 1999).

The most likely scenario to explain these phenomena consists of an
ionized region around the central engine, surrounded by an accretion
disk or torus with a scale height of at least 30 pc which is comprised
primarily of atomic gas.  Assuming a radius of $\sim$50 pc based on
the extent of the absorption and the orientation of the jet axis
(between 65$-$80$^{\circ}$ to the line of sight) we can estimate, from rough
virial arguments, an enclosed mass of M$\sim$5x10$^8$ M$_\odot$.




\end{document}